# A Densely Interconnected Network for Deep Learning Accelerated MRI

Jon Andre Ottesen[1,2]

Matthan W.A. Caan[1,3]

Inge Rasmus Groote[1,4]

Atle Bjørnerud[1,2]

[1]Computational Radiology & Artificial Intelligence (CRAI) Unit, Division of Radiology and Nuclear Medicine, Oslo University Hospital, Oslo, Norway

[2]Department of Physics, Faculty of Mathematics and Natural Sciences, University of Oslo, Oslo, Norway

[3] Amsterdam UMC, University of Amsterdam, Biomedical Engineering and Physics, Amsterdam, Netherlands.

[4]Department of Radiology, Vestfold Hospital Trust, Tønsberg, Norway

Corresponding Author: Jon Andre Ottesen

**Acknowledgments:**

The project was supported by the Norwegian South-Eastern Health Authority (grant number 2021031). In addition, the authors would like to thank Dr. Endre Grøvik for carefully reading the manuscript and the University of Oslo for access to the computation resources used for the project.

# Abstract

**Objective:** To improve accelerated MRI reconstruction through a densely connected cascading deep learning reconstruction framework.

**Materials and Methods:** A cascading deep learning reconstruction framework (baseline model) was modified by applying three architectural modifications: Input-level dense connections between cascade inputs and outputs, an improved deep learning sub-network, and long-range skip-connections between subsequent deep learning networks. An ablation study was performed, where five model configurations were trained on the NYU fastMRI neuro dataset with an end-to-end scheme conjunct on four- and eight-fold acceleration. The trained models were evaluated by comparing their respective structural similarity index measure (SSIM), normalized mean square error (NMSE) and peak signal to noise ratio (PSNR).

**Results:** The proposed densely interconnected residual cascading network (DIRCN), utilizing all three suggested modifications, achieved a SSIM improvement of 8% and 11% for four- and eight-fold acceleration, respectively. For eight-fold acceleration, the model achieved a 23% decrease in the NMSE when compared to the baseline model. In an ablation study, the individual architectural modifications all contributed to this improvement, by reducing the SSIM and NMSE with approximately 3% and 5% for four-fold acceleration, respectively.

**Conclusion:** The proposed architectural modifications allow for simple adjustments on an already existing cascading framework to further improve the resulting reconstructions.

**Keywords:** MRI, Deep learning, Image reconstruction

# Introduction

Magnetic resonance imaging (MRI) data acquisition is an inherently slow process due to fundamental physical constraints that limit the rate of k-space traversal. This can lead to prolonged MRI sequences, during which the patient must remain still to achieve images of diagnostic quality. Traditionally, parallel imaging [1–3] and compressed sensing [4] have been used to reduce aliasing artifacts caused by the subsampling of k-space. This allows for the reconstruction of clinically acceptable images with up to two-fold acceleration for brain MRI [5].

In recent years, deep learning and convolutional neural networks (CNNs) have shown great promise as an alternative framework for MRI reconstruction to further accelerate scans beyond that of parallel imaging and compressed sensing. A study has shown that the end-to-end variational network [6] can reconstruct images that are interchangeable for the detection of internal derangements of the knee when compared to their fully sampled counterparts at four-fold acceleration [7].

Deep learning MRI reconstruction frameworks span a wide variety of different architectures, from U-Net-based models [8] with [9] and without [10] data consistency in k-space; general adversarial networks [11]; k-space reconstruction networks [12]; and cascaded networks that consisting of sub-networks [6, 13–17], in which the latter performs excellently [18, 19]. The sub-networks perform reconstruction in the image domain, frequency domain or both, and reconstruct the complete image or frequency information from the subsampled scan. CNNs are commonly used as sub-networks, and the architectures range from shallow sub-networks tallying a few convolutional layers per cascade [15, 20], to deeper architectures [6, 13, 21].

In this work, we sought to improve the overall reconstruction quality for cascading networks, by introducing and testing three novel improvements. The end-to-end variational network [6] was adopted as a baseline model, and from this we developed the densely interconnected residual cascading network (DIRCN). The contributions of DIRCN are summarized as follows:

1. Input level dense connections [22] were implemented to improve gradient and information flow through cascades in a similar manner to previous implementations [15, 23].
2. A U-Net-based sub-network that incorporates aggregated [24] residual connections [25] with squeeze-and-excitation [26], and the sigmoid linear unit (SiLU) activation function [27–29] was adopted to improve in-cascade gradient flow and expressivity through channel-wise excitation.

3. Long range skip-connections across sub-networks were implemented; we hypothesized that long range skip-connections will further improve gradient flow and further fine-tune feature maps.

Focus was placed on facilitating gradient flow and connectivity between sub-networks.

The architectural modifications proposed in this study were tested on the NYU fastMRI neuro dataset [30, 31] to gauge the importance of the input-level dense connections, long range skip-connections, and the proposed U-Net-based sub-network for four- and eight-fold k-space subsampling.

## Methods

This section provides an overview of the problem formulation, network architecture, dataset, training scheme and model evaluation. Additional details regarding the model implementation are given in the source repository.[1]

*Problem formulation*

For 2D cartesian acquisition, let $k \in \mathbb{C}^{c \times n_{kx} \times n_{ky}}$ denote the fully sampled multi-coil complex-valued k-space representation for $c$ receiver coils with $n_{kx}$ and $n_{ky}$ sampled datapoints along the frequency and phase encoding dimensions, respectively. The corresponding image representation $x \in \mathbb{C}^{c \times n_{kx} \times n_{ky}}$ of the sampled k-space for the j-th coil element is related by

$$k_j = \mathcal{F}(S_j \circ x_j) + \epsilon, \qquad \text{Eq. 1}$$

where $\mathcal{F}$ is the two-dimensional Fourier transform, $S_j$ is the coil sensitivity for the j-th receiver coil, $\circ$ is the Hadamard product (element-wise multiplication) and $\epsilon$ is additive noise.

The speed by which k-space is traversed is governed by the number of phase encoding steps $n_{ky}$. To accelerate MRI acquisition, k-space can be subsampled by reducing the number of phase-encoding steps. From the fully sampled k-space data, $k$, the subsampled subset of k-space is given by

$$k_u = U \circ k, \qquad \text{Eq. 2}$$

where $k_u \in \mathbb{C}^{c \times n_{kx} \times n_{ky}}$ is the undersampled k-space and $U \in \mathbb{C}^{n_{kx} \times n_{ky}}$ is a binary undersampling mask. The acceleration factor is the ratio between the number of masked lines and the total number of acquired lines.

---

[1]https://github.com/JonOttesen/DIRCN

The intention of image reconstruction is to solve the inverse problem of recovering the image representation $x$ from the undersampled k-space, $k_u$. To that end, supervised deep learning networks aim to map a subsampled k-space to the corresponding fully sampled k-space by learning from pairs of undersampled and fully sampled scans.

*Network architecture*

This work presents a densely interconnected residual cascading network (DIRCN) for MRI reconstruction. DIRCN builds on top of the end-to-end variational network [6], and the end-to-end variational network was adopted as the baseline model. The novelty in this work stems from the three extensions employed to this baseline model. With these modifications we sought to improve the gradient flow and connectivity between the cascading layers. To that end, the baseline model was extended by: (1) Long range dense input-level connections, (2) a U-Net-based CNN sub-network and (3) long-range skip connections dubbed interconnections. The general DIRCN model architecture is illustrated in Fig. 1, where the U-Net-based CNN sub-network is replaced by a simplified CNN for readability.

*Baseline model*

The baseline network follows the end-to-end variational network [6], but the variational update mechanism implemented by Hammernik et al. [32] was changed to a data consistency method similar to Schlemper et al. [15]. Given a set of subsampled k-space, $k_u$, and a corresponding k-space prediction, $k_p$, data consistency was be implemented by

$$f_{dc}(k_p) = \begin{cases} \frac{k_u + \lambda \cdot k_p}{1+\lambda} & U_{i,j} = 1 \\ k_p & U_{i,j} = 0 \end{cases}, \qquad Eq.\ 3$$

where $\lambda$ is a learnable parameter initialized to 0.01.

Given an undersampled k-space sample, $k_u$, the coil sensitivities, $S$, were estimated identical to the baseline model with a CNN using the fully sampled center portion of k-space. Note that the network architecture used for coil sensitivity estimation was identical to the sub-networks embedded for image reconstruction, except a lower number of parameters to reduce memory constraints.

The baseline model consists of $m$ cascades, each of which consists of a series of four distinct operations:

1. The coil dimensionality is reduced by $I_{red} = \sum_{i=1}^{n_c} \mathcal{F}^{-1}(k_u^i)\overline{S_i}$,[2] effectively reducing the number of channels from c-coils to a single complex image.

---

[2] Overline denotes the complex conjugate, not complex transpose.

2. The coil reduced image is refined by a CNN: $I_{ref} = CNN(I_{red})$, where $I_{ref}$ is the refined complex coil reduced image.
3. The number of coils in the refined image is expanded back to the original number using the coil sensitivities by $I_{ep} = cat(I_{ref} \circ S_1, I_{ref} \circ S_2, \ldots, I_{ref} \circ S_{n_c-1}, I_{ref} \circ S_{n_c})$, where the *cat* operation is concatenation along the channel dimension.
4. Data consistency is enforced, and the data consistent k-space is given by $k_{dc} = f_{dc}(\mathcal{F}(I_{ep}))$, where $f_{dc}$ is given in Eq. 3.

The cascade output, $k_{dc}$, was used as the input for the next cascade. In this work, the number of cascades was enforced to $m = 12$ for all model configurations. The magnitude image was computed by taking the root sum of squares (RSS) and subsequently the complex absolute.

*Input-level dense connections*

As a first extension, input level dense connections [22] were implemented to facilitate gradient and information flow throughout the network. For the k'th cascade, the CNN refinement from step 2 can be written as $I_{ref}^k = CNN^k(I_{red}^k)$.[3] In the case of input-level dense connections, the CNN input is given by the concatenated coil reduced image from the prior cascades. The CNN refinement step for input level dense connections is given by

$$I_{ref}^k = CNN^k\left(cat(I_{red}^k, I_{red}^{k-1}, \ldots, I_{red}^2, I_{red}^1)\right), \qquad \text{Eq. 4}$$

and the input-level dense connections are illustrated in Fig. 1.

*Refinement of CNNs: ResXUnet*

The second extension originates from the multiple alterations and refinements which have been proposed based on the U-Net [33]. These U-Net alterations utilize different architectural modifications such as residual connections, dense connections, attention mechanisms and multilayer feature fusion, amongst others. In this work, a modified U-Net-based model dubbed ResXUNet was embedded into the cascaded network, incorporating aggregated residual connections, squeeze-and-excitation and the SiLU activation function. Residual connections facilitate gradient flow [25], squeeze-and-excitation model channel-wise dependencies [26] and the SiLU activation function has shown improved performance over other activation functions [27–29]. The ResXUNet model is illustrated in Fig. 2.

*Long range skip connections – interconnections*

---

[3] Superscript denote the cascade number.

The cascading network type can be seen as a series of independent sub-networks, where the input of a sub-network is the data consistent output from a prior sub-network. Besides this connection, each individual sub-network does not share any of the extracted feature maps from a prior sub-network.

To improve the interconnectivity between sub-modules, the third extension is to insert long-range skip connections comparable with those utilized in the U-Net. The interconnections were implemented to connect every subsequent sub-model, thereby creating a flow of feature maps between the sub-networks. This was done by copying the final feature map for each resolution in the deep learning model and concatenating the feature maps for each resolution onto the subsequent sub-network. These connections, coined interconnections, are illustrated in Fig 1.

*Dataset and undersampling masks*

The proposed method was trained, validated and evaluated on the fully sampled raw k-space fastMRI neuro dataset [30, 31]. The dataset consists of two predetermined splits, one for training with 4469 scans and one for validation with 1378 scans. Both sets consist of T1-weighted pre and post contrast, T2-weighted and FLAIR images from both 1.5T and 3T scanners. The scans have a wide variety of acquisition matrices with and without zero-padding. The predetermined validation set was randomly split up into a test and validation set, with 689 scans in both the validation and test set. The exact distribution used can be found in the source repository.[1]

The fully sampled raw k-space was undersampled by an equidistant downsampling scheme with a fully sampled center. This scheme was used for both four- and eight-fold acceleration, and the center contained 8% or 4% of the number of phase-encoding steps, respectively.

To reduce memory requirements, frequency oversampling was removed from the data. This was done by quadratically cropping all images in the image domain, followed by Fourier transforming them back into k-space before being undersampled. The ground truth image is the fully sampled RSS combined complex absolute image cropped quadratically. The preprocessing steps are illustrated in Fig. 3.

*Implementation*

In total five model configurations were trained. This includes the baseline model and DIRCN, after which an ablation study where the three architectural modifications: input-level dense connections, interconnections, and ResXUNet were tested individually.

All model configurations were trained and implemented in Python using PyTorch version 1.7.1 [34]. The Adam optimizer [35] was used with default PyTorch parameters and an initial learning rate of 0.002, with stepwise learning-rate decay every 60'th iteration using $\gamma = 0.1$ and Amsgrad [36] enabled. All models were trained for 120 iterations, with a mini-batch size of one, and every iteration looped over 10,000 randomly selected image slices from the dedicated NYU fastMRI neuro training set. After each iteration, the models were validated on 4,000 randomly selected image slices from the validation set. Each image was undersampled with equal likelihood by either four- or eight-fold acceleration during training and validation. Neither data augmentation nor data parallelization was used. The number of parameters was set to approximately 45 million for all model configurations to constrain memory usage for the most memory-intensive models.

Training took approximately ten days for all models not using the ResXUNet model and 20 days for the models that used the ResXUNet architecture. All training was done on either a Nvidia V100 (32 GB) or a RTX 3090 (24 GB). All networks were benchmarked on a single RTX 2080 Ti (11 GB). The inference time was computed as the mean of 1000 reconstructions on a single four-fold accelerated slice of size $376 \times 376$ with 20 coil elements and randomly initialized model parameters.

The loss function was an equally weighted linear combination of the Gaussian weighted SSIM and the mean absolute distance (L1 loss). Reconstruction quality was assessed using the normalized mean square error (NMSE), peak signal to noise ratio (PSNR) and structural similarity index measure (SSIM) [37]. All model configurations were evaluated using the final checkpoint after 120 iterations.

## Results

Violin plots of the SSIM-values for the baseline model and DIRCN on the test set are shown in Fig. 4. An improved mean SSIM can be observed for all weighting schemes, this effect is more pronounced for eight-fold acceleration compared to four-fold acceleration.

The mean SSIM, NMSE and PSNR for the four- and eight-fold accelerated images for the test dataset for the different model configurations are given in Table 1. We see an approximate improvement of 2-4% for the SSIM metric for all suggested modifications for both acceleration factors. The corresponding NMSE improvement is more pronounced, with an approximate improvement of 10% for all suggested modifications over the baseline model for eight-fold acceleration.

DIRCN achieved superior PSNR, NMSE and SSIM compared to the baseline model for both four- and eight-fold acceleration. DIRCN achieved an improvement in the SSIM, NMSE and PSNR that is close to the additive individual improvement for the dense connections, interconnections, and ResXUNet model, individually. For training, the memory consumption for the different models were approximately 15 GB for the baseline model; 15.4 GB for the input-level dense connected model; 16 GB for the interconnected model; 30 GB for the ResXUNet model; and 30 GB for DIRCN.

Figures 5, 6 and 7 show representative reconstructions of magnitude T1-weighted, T2-weighted and FLAIR images with their respective absolute error for the baseline and DIRCN. A visual decrease in the absolute error between the baseline model and DIRCN can be observed, especially for the eight-fold accelerated images. Typically, DIRCN produces reconstructions that are closer to that of the ground truth image, as can be seen from the error maps. One such example can be discerned from the error map of the eight-fold accelerated T2-weighted images. For the eight-fold accelerated images, DIRCN show visible artifacts, but still outperforms the baseline model for the same acceleration.

The training and validation losses for the baseline model and DIRCN are plotted in Fig. 8, and the validation loss for all network configurations are plotted in Fig. 9. A difference in convergence can be seen across the different configurations, and the dense and residual configurations have a high initial convergence. The configuration with interconnections had a similar initial convergence to that of the baseline model. However, the convergence rate increased after an initial phase. No major sign of overfitting can be discerned from Fig. 8; there is, however, a slight divergence between the training and validation loss. This divergence was observed for all model configurations, but slightly more pronounced for DIRCN when compared to the baseline model.

## Discussion

The DIRCN showed superior metrical results compared to the baseline model, with the error maps being closer to that of the ground truth image, and the SSIM and NMSE showing over 10% increase for both acceleration factors. Although the reconstructions are closer to that of the ground truth, it is difficult to discern any visual difference on the magnitude images, since the reconstructions of the baseline model are state-of-the-art.

Similar to a previous work [16], this work showed an improved performance with the addition of input-level dense connections. In addition, the dense connections had no noticeable inference time or memory overhead when compared to the baseline model. The proposed interconnections showed a similar increase in performance as the dense connections, with no noticeable increase in inference time or memory overhead. The minor increase in the number of parameters came from an increase in the number of incoming channels in the concatenation operation. Unlike the dense connections, the implementation of the interconnections can be modified within the network to further improve performance. A possible improvement could be to use attention similar to attention U-Net [38]. In addition, interconnections could be implemented through dense connections, which may further improve the overall performance. However, this was opted against to avoid additional computational bottlenecks, and was outside the scope of this study.

The ResXUNet model used in this study achieved improved performance in comparison to the baseline model. However, this improved performance introduces additional computational complexity, which translates to increased memory consumption and inference time. Because of these overheads, additional research into the most suitable sub-network architecture is necessary. Nonetheless, the increase in performance may warrant the additional computational complexity. Future studies should be performed to find an ideal tradeoff between computational overhead and overall performance increase.

The results shown in this work may further improve through more optimized training strategies, such as parallelization, data augmentation or a better optimizer. Additionally, separate training for four- and eight-fold acceleration and extended training time could further improve the results.

The study has limitations in that the model has only been trained on retrospective public domain data. As such, it is necessary to further test the model on clinically valid prospective data on in-house MRI systems. In addition, in this work the effects of the undersampling scheme on the model extensions were not evaluated. However, as the enhancements are of architectural nature, it is not unreasonable to assume that other undersampling schemes may benefit from the proposed enhancements.

## Conclusion

Inspired by the end-to-end variational network, multiple architectural improvements were tested and evaluated. Experimental result demonstrates that input-level dense connections, a modified convolutional sub-network and interconnections (long-range skip connections) improved the quality of the reconstructed images for both four- and eight-fold acceleration. Our findings suggest the importance of gradient flow and shared information between cascades for MRI reconstruction networks. The proposed DIRCN attains improved results over the baseline model, and more fine structures were visibly preserved for eight-fold acceleration in the reconstructions. It is shown that simple alterations and additions to enhancing the cascading framework can improve the overall quality of the reconstruction.

## Author contributions

Study conception and design: all authors; acquisition of data, data preprocessing and model training: JAO; analysis and interpretation of data: all authors; drafting of manuscript. all authors; critical revision: all authors.

## Conflict of Interest

M.W.A. Caan is shareholder of Nico.lab International Ltd.

Table 1: The structural similarity index measure (SSIM), normalized mean square error (NMSE) and peak signal to noise ratio (PSNR) for the different architectural modifications for T1-weighted, T2-weighted, FLAIR and ALL images from the dedicated test dataset for four-fold and eight-fold acceleration. In addition, the number of parameters and inference time is given for each model configuration.

| Architecture | #Params | Inference time [ms] | Modality | Four-fold acceleration | | | Eight-fold acceleration | | |
|---|---|---|---|---|---|---|---|---|---|
| | | | | SSIM | NMSE | PSNR | SSIM | NMSE | PSNR |
| Baseline | 45M | 148 ± 3 | T1 | 0.9626 | 0.0033 | 41.5 | 0.9466 | 0.0070 | 38.2 |
| | | | T2 | 0.9556 | 0.0044 | 40.0 | 0.9399 | 0.0096 | 36.6 |
| | | | FLAIR | 0.9357 | 0.0055 | 39.1 | 0.9123 | 0.0111 | 36.1 |
| | | | ALL | 0.9560 | 0.0041 | 40.4 | 0.9395 | 0.0088 | 37.0 |
| Baseline + Dense | 45M | 153 ± 2 | T1 | 0.9637 | 0.0031 | 41.7 | 0.9486 | 0.0064 | 38.6 |
| | | | T2 | 0.9569 | 0.0042 | 40.2 | 0.9420 | 0.0085 | 37.0 |
| | | | FLAIR | 0.9378 | 0.0053 | 39.3 | 0.9154 | 0.0100 | 36.5 |
| | | | ALL | 0.9574 | 0.0039 | 40.6 | 0.9417 | 0.0080 | 37.5 |
| Baseline + ResXUNet | 41M | 394 ± 4 | T1 | 0.9635 | 0.0031 | 41.7 | 0.9485 | 0.0065 | 38.6 |
| | | | T2 | 0.9565 | 0.0042 | 40.2 | 0.9417 | 0.0087 | 36.9 |
| | | | FLAIR | 0.9370 | 0.0053 | 39.3 | 0.9149 | 0.0101 | 36.5 |
| | | | ALL | 0.9569 | 0.0040 | 40.6 | 0.9415 | 0.0081 | 37.4 |
| Baseline + Interconnections | 49M | 159 ± 2 | T1 | 0.9638 | 0.0030 | 41.8 | 0.9482 | 0.0065 | 38.6 |
| | | | T2 | 0.9567 | 0.0041 | 40.3 | 0.9419 | 0.0085 | 37.0 |
| | | | FLAIR | 0.9375 | 0.0053 | 39.3 | 0.9144 | 0.0101 | 36.5 |
| | | | ALL | 0.9573 | 0.0039 | 40.7 | 0.9415 | 0.0080 | 37.5 |
| DIRCN | 47 M | 387 ± 5 | T1 | 0.9658 | 0.0028 | 42.2 | 0.9529 | 0.0054 | 39.3 |
| | | | T2 | 0.9588 | 0.0038 | 40.7 | 0.9460 | 0.0072 | 37.7 |
| | | | FLAIR | 0.9408 | 0.0048 | 39.8 | 0.9216 | 0.0085 | 37.2 |
| | | | **ALL** | **0.9594** | **0.0035** | **41.1** | **0.9460** | **0.0068** | **38.2** |

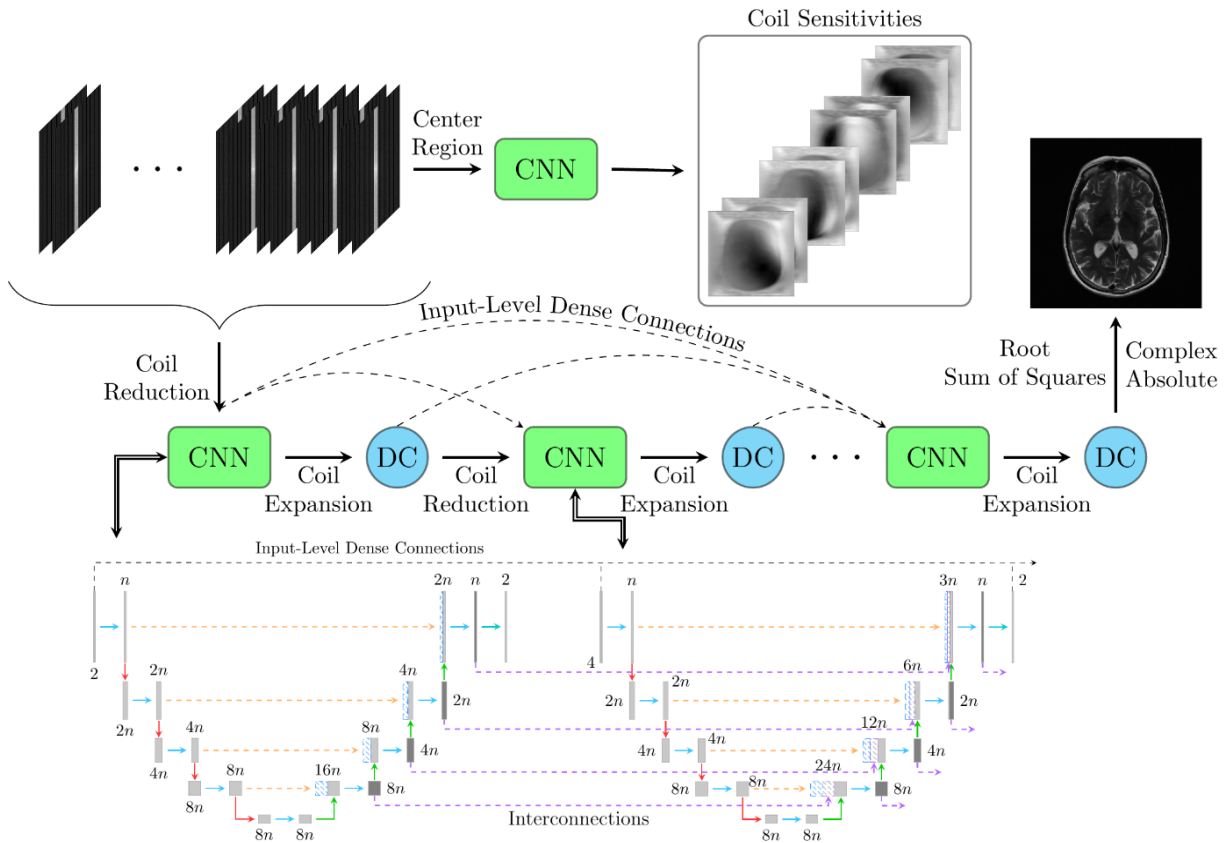

*Fig. 1* An illustration of the dense interconnected residual cascading network (DIRCN). The model consists of m cascades of a simplified U-Net-based architecture and data consistency (DC). Each cascade is connected to every prior cascade by input-

*level dense connections illustrated by the black dashed lines. Every sub-network is connected to the prior sub-network through concatenation, dubbed interconnections and they are illustrated by the purple dashed lines. The output is the root-sum-of-squares image of the data-consistent output from the last cascade.*

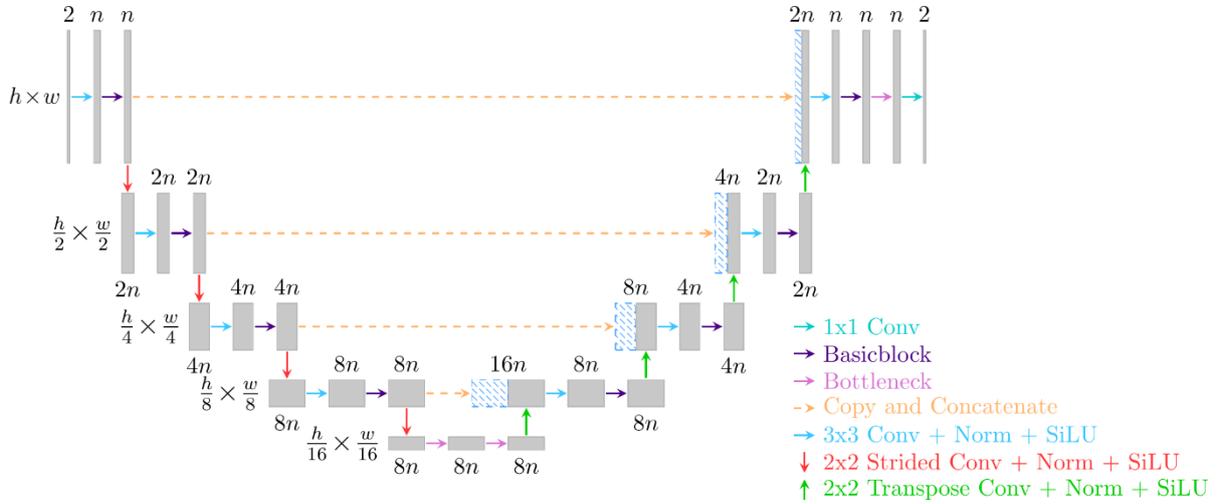

***Fig.* 2** *The suggested ResXUNet model used for the CNN refinement step. The model includes aggregated residual connections* [24] *for improved gradient flow, squeeze-and-excitation* [26] *for learnable channel wise attention and the SiLU activation function* [27–29]. *The squeeze-and-excitation operation is implemented through the residual blocks.*

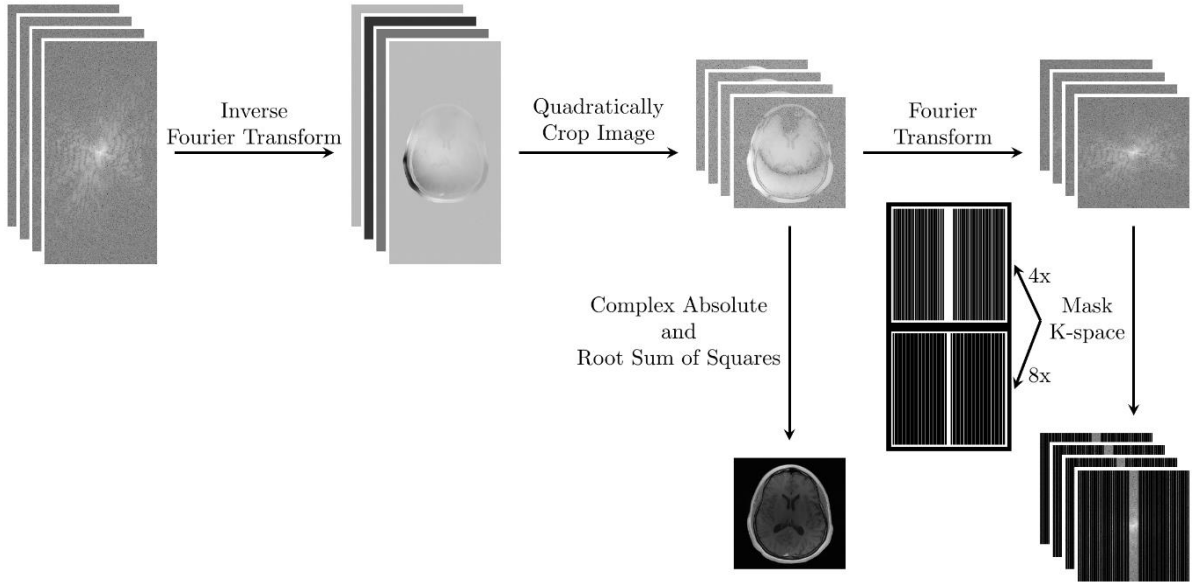

***Fig.* 3** *An illustration of the preprocessing steps for the undersampled k-space and the fully sampled magnitude image. Raw multi-coil k-space data was first Fourier transformed to image space, then quadratically cropped along the height and width dimension for all coils. The ground truth image was the complex absolute and root sum of squares of the complex coil images. For model inputs, the cropped complex coil images were Fourier transformed back to k-space before being masked by either a four- or eight-fold downsampling mask.*

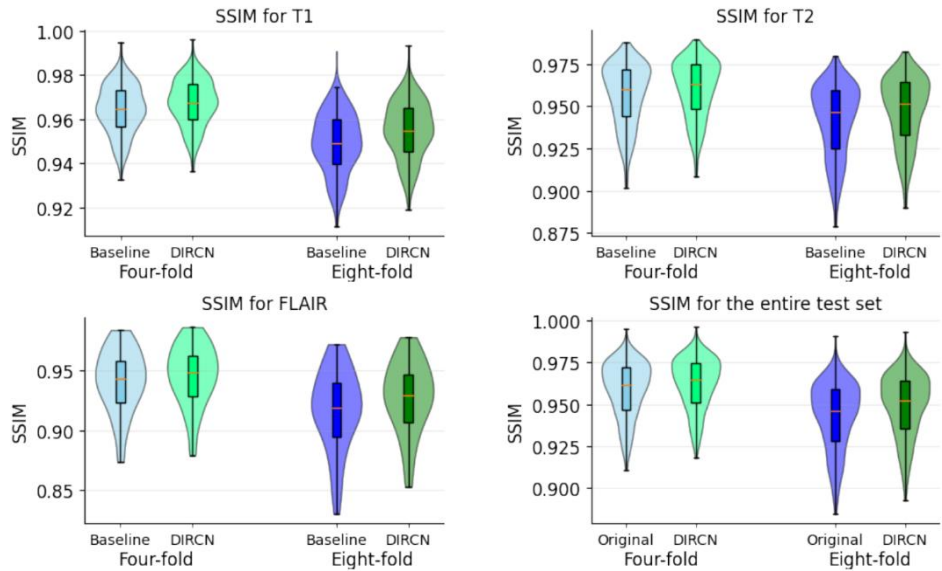

*Fig. 4 Structural similarity index measure (SSIM) distributions for the designated test dataset for baseline model and DIRCN for T1-weighted, T2-weighted, FLAIR and all images. The distributions show the SSIM for both four- and eight-fold acceleration. Note, alle outlier for low SSIM values are emitted for readability, with the outlier definition following regular conventions.*

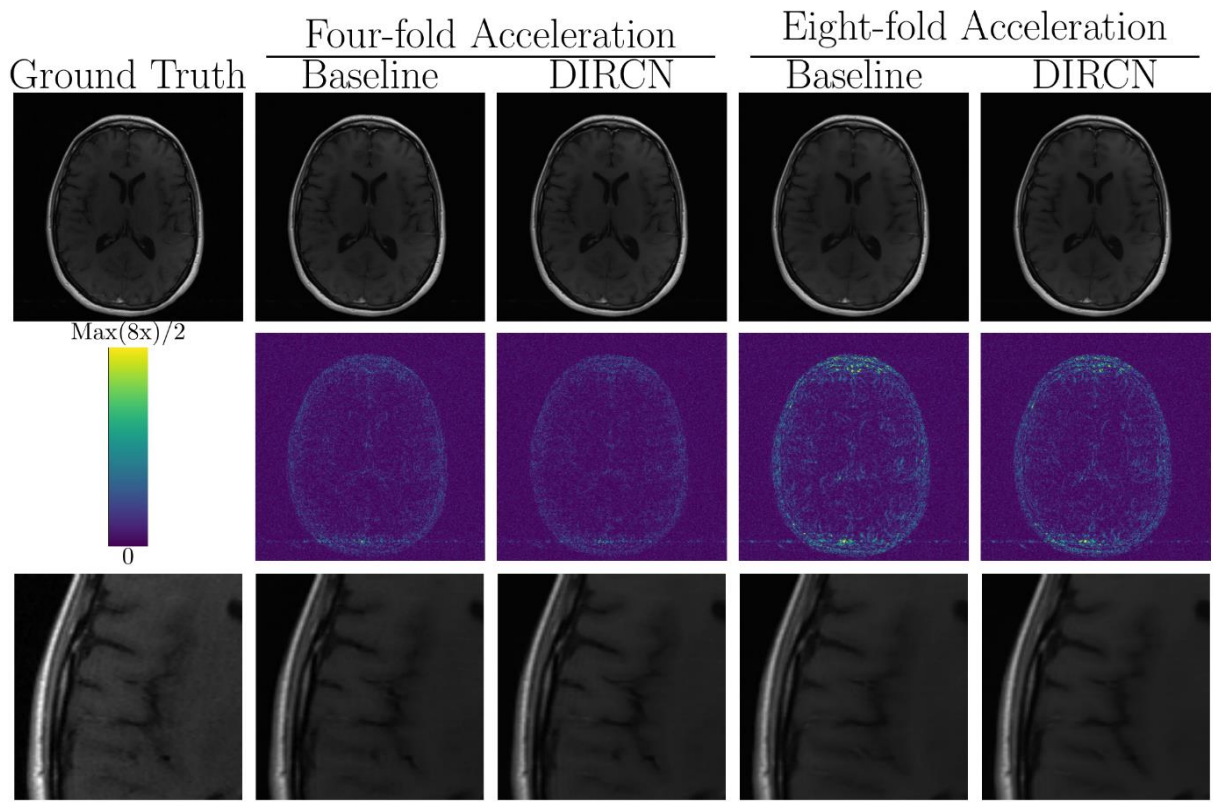

*Fig. 5 A representative example of a T1-weighted reconstruction with the baseline model and DIRCN for four- and eight-fold acceleration. This includes their respective reconstructions and the corresponding error map (absolute difference) between the fully sampled image and the reconstruction. The colormap goes between 0 and half the maximum error for eight-fold*

*acceleration to emphasize visual difference. The bottom images are a region of interest where slight improvement between the baseline and DIRCN can be seen at close inspection.*

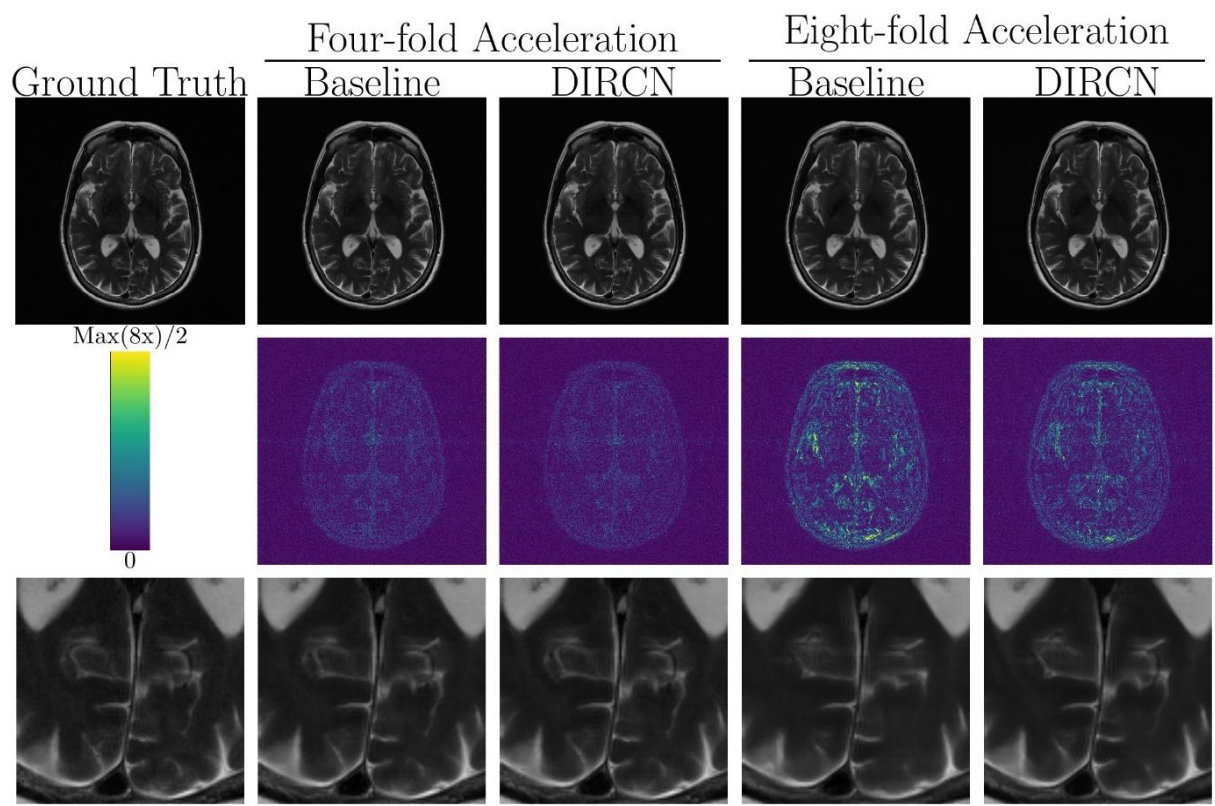

**Fig. 6** *A representative example of a T2-weighted reconstruction with the baseline model and DIRCN for four- and eight-fold acceleration. This includes their respective reconstructions and the corresponding error map (absolute difference) between the fully sampled image and the reconstruction. The colormap goes between 0 and half the maximum error for eight-fold acceleration to emphasize visual difference. The bottom images are a region of interest where differences between the baseline model and DIRCN reconstructions for eight-fold acceleration can be seen.*

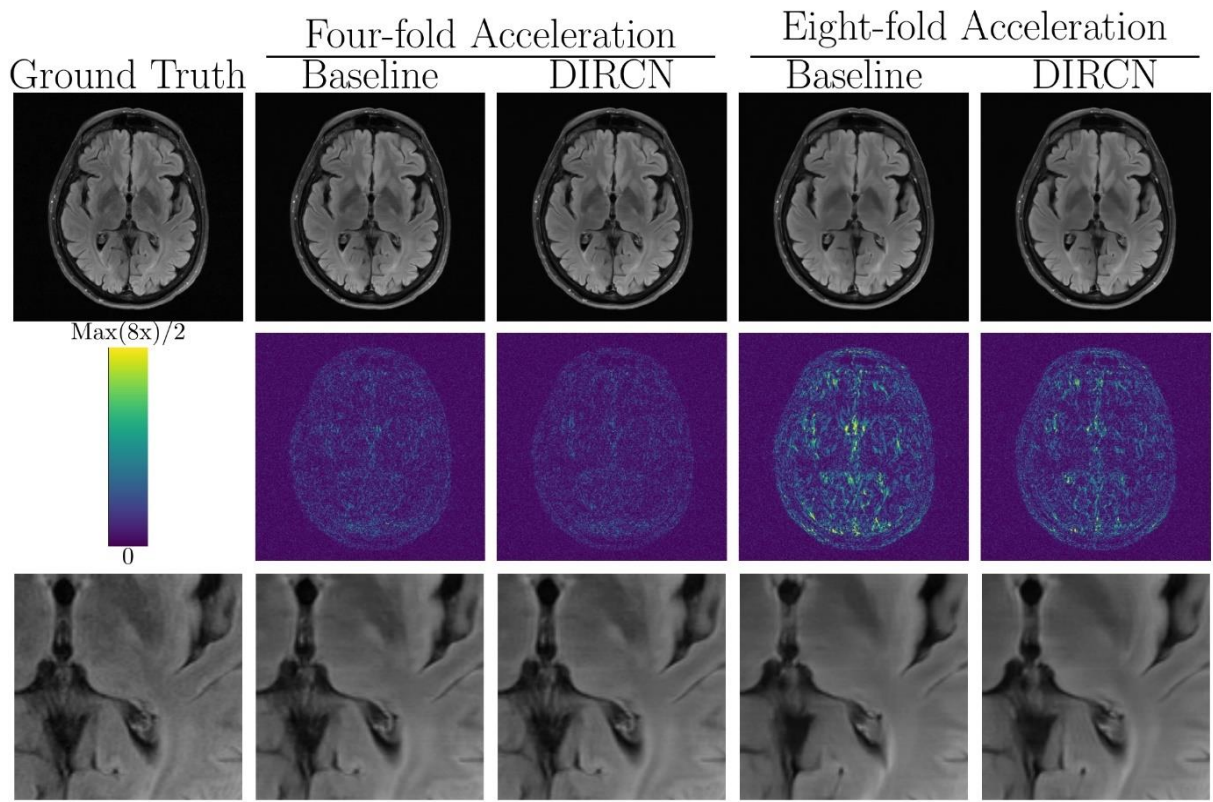

**Fig. 7** A representative example of a FLAIR reconstruction with the baseline model and DIRCN for four- and eight-fold acceleration. This includes their respective reconstructions and the corresponding error map (absolute difference) between the fully sampled image and the reconstruction. The colormap goes between 0 and half the maximum error for eight-fold acceleration to emphasize visual difference. The bottom images are a region of interest where one can see an erroneous reconstruction for eight-fold acceleration for both models.

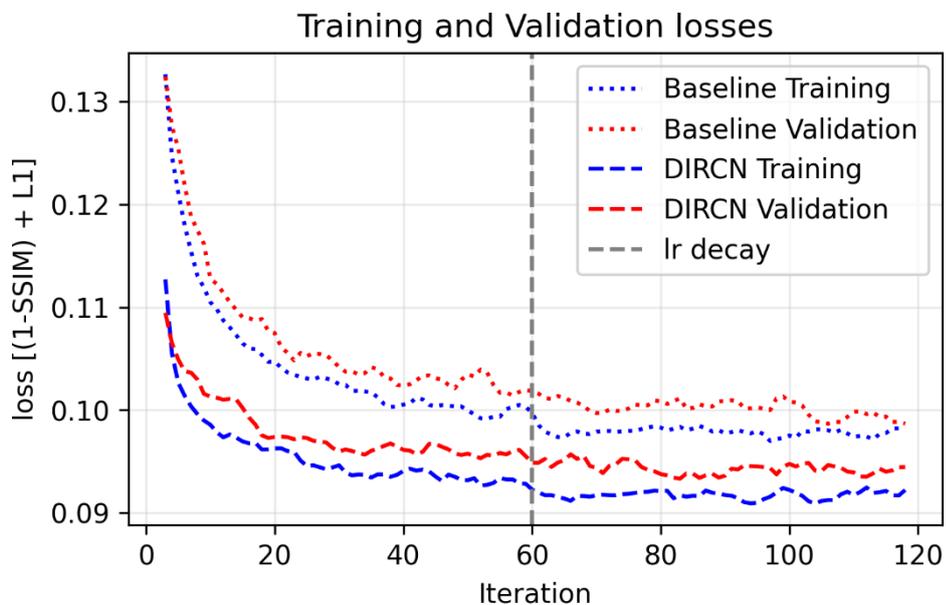

**Fig. 8** The training and validation losses for the baseline and DIRCN for 120 iterations. The plotted losses are mean losses for a 5-point sliding window starting at iteration 3 and ending at iteration 117.

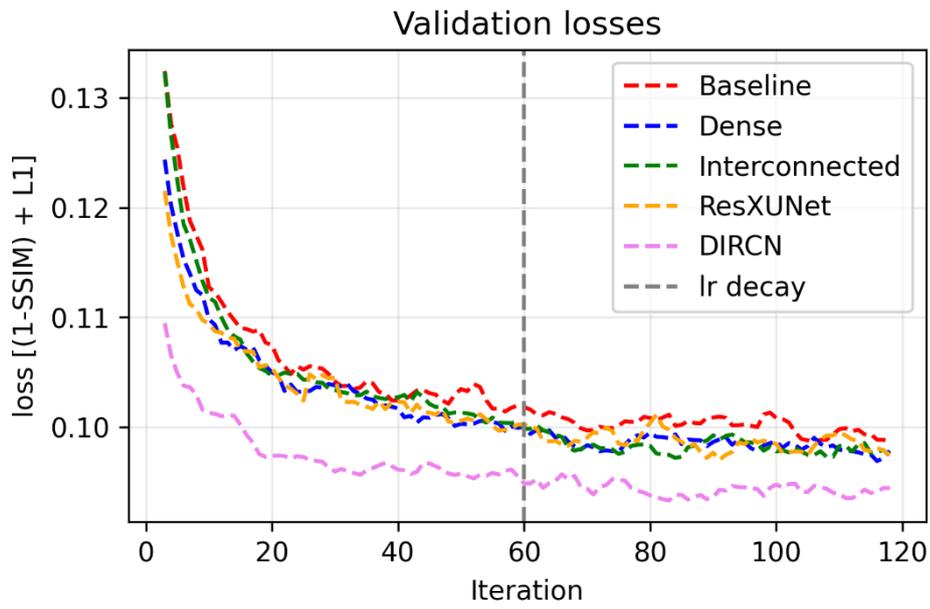

*Fig.* 9 *The validation losses for all network configurations for 120 iterations. The plotted losses are mean losses for a 5-point sliding window starting at iteration 3 and ending at iteration 117.*